\documentclass[twocolumn,english,reprint,prd,aps,superscriptaddress,nofootinbib,floatfix]{revtex4-1}
\usepackage[latin9]{inputenc}
\usepackage{babel}
\usepackage{bm}
\usepackage{amsmath}
\usepackage{amssymb}
\usepackage[unicode=true,
 bookmarks=false,
 breaklinks=false,pdfborder={0 0 1},backref=section,colorlinks=false]
 {hyperref}

\makeatletter
\usepackage{babel}

\usepackage{graphicx}
\usepackage{bm}\usepackage{color}\usepackage{amsfonts}
\usepackage{footmisc}

\makeatother

\begin{document}
\title{Infrared scaling for a graviton condensate }
\author{Sougato Bose}
\affiliation{Department of Physics and Astronomy, University College London, Gower
Street, WC1E 6BT London, UK}
\author{Anupam Mazumdar}
\affiliation{Van Swinderen Institute, University of Groningen,Groningen, 9747 AG,
The Netherlands}
\author{Marko Toro\v{s}}
\affiliation{School of Physics and Astronomy, University of Glasgow, Glasgow, G12
8QQ, UK}
\begin{abstract}
The coupling between gravity and matter provides an intriguing length
scale in the infrared for theories of gravity within Einstein-Hilbert
action and beyond. In particular, we will show that such an infrared
length scale is determined by the number of gravitons $N_{g}\gg1$
associated to a given mass in the non-relativistic limit. After tracing
out the matter degrees of freedom, the graviton vacuum is found to
be in a displaced vacuum with an occupation number of gravitons $N_{g}\gg1$.
In the infrared, the length scale appears to be $L=\sqrt{N_{g}}\ell_{p}$,
where $L$ is the new infrared length scale, and $\ell_{p}$ is the
Planck length. In a specific example, we have found that the infrared
length scale is greater than the Schwarzschild radius for a slowly
moving in-falling thin shell of matter. We will argue that the appearance
of such an infrared length scale in higher curvature theories of gravity,
such as in quadratic and cubic curvature theories of gravity, is also
expected. Furthermore, we will show that gravity is fundamentally
different from the electromagnetic interaction where the number of
photons, $N_{p}$, is the \textit{fine structure constant} after tracing
out an electron wave function. 
\end{abstract}
\maketitle

\section{Introduction}

It is believed that the gravitational interaction is mediated by the
spin-2 graviton, which can be canonically quantised around a weak
curvature background \citep{Gupta}. A massless graviton in four spacetime
dimensions will have both 2-on-shell and 6-off-shell degrees of freedom~\citep{dof}.
The former is responsible for describing independent dynamical modes
such as gravitational waves, while the latter describes how the force
is being mediated between the two masses. Despite all our efforts
the discovery of a graviton remains a challenging problem, see \citep{Baym,Dyson,Ashoorioon:2012kh}.
However, the quantum nature of a graviton leaves indelible mark in
both classical and quantum systems~\citep{Ng,Amelino,Parikh,Toros:2020dbf,Bose:2017nin,Vedral,Marshman:2019sne}.

Despite of the weakness in the gravitational interaction, gravity
is unique among the other known fundamental interactions of nature
that it generates a new length scale in the presence of a self-gravitating
matter~\citep{Ruffini}. The gravitational radius for a given non-rotating
mass, $M$, is given by $r_{g}=2GM$, which is known as the Schwarzschild
radius or the gravitational radius, while the Planck length, which
is determined solely by fundamental constants~\footnote{We are working in natural units $c=\hbar=1$, and $\epsilon_{0}=1$.
The metric signature is given by $(-,+,+,+)$ and Einstein summation
convetion will be used in the text.}, is given by $\ell_{p}=\sqrt{G}$. In Ref.~\citep{Dvali2013black},
the idea has been proposed from the corpuscular nature of a black
hole -- a black hole is a condensate of gravitons~\citep{Dvali2013black,Dvali:2015,Casadio:2015,Dvali:2021},
whose occupation number can be denoted here by $N_{g}\sim(M/M_{p})^{2}\gg1$
for any mass $M\geq M_{p}$. The large occupancy leads to not only
weakening of the gravitational strength by $\sim1/\sqrt{N_{g}}$,
but also leads to classicalization of a black hole, therefore, recovers
the classical black hole spacetime geometry outside the Schwarzschild
radius. In this regard, we might imagine that $N_{g}$ would dictate
how classical the space time of a black hole would appear to be for
a far away observer. Thus the new length scale appears in the infrared;
$L\sim\sqrt{N_{g}}/M_{p}\gg\ell_{p}$ for $N_{g}\gg1$.

Recently, a very similar result were obtained by us in a quantum system~\citep{Bose:2021ekc},
where both matter and gravity were treated as a quantum entity in
a perturbative regime. We found that by \textit{tracing out} the non-relativistic
self-gravitating matter of mass $M$, the graviton vacuum state is
found to be that of a displaced vacuum, like a coherent state with
the occupation number similar to that of $N_{g}\sim(M/M_{p})^{2}$.
For $N_{g}\gg1$, the gravitons can be thought of as a condensate
of mass $M$. For a light subatomic particle, such as that of an electron,
the number of gravitons by tracing out the electron is much less than
unity, $N_{g}\sim(m_{e}/M_{p})^{2}\sim10^{-44}\ll1$~\footnote{Interestingly, the electron cannot be described by a gravitational
metric, such as a Reissner-Nordström or a Kerr metric \citep{Burinskii:2005mm}.
The metric is inherently a classical notion.}~\footnote{There is another proposal, known as the fuzz ball paradigm \citep{Mathur:2005zp},
where it has been argued that the new scale in gravity will arise
naturally from the quantum fluctuations in the gravitational degrees
of freedom~\citep{Mathur:2020ely}, in particular by taking all microscopic
states of string theory, namely the fuzz ball states~\citep{Mathur:2005zp}.
The fuzz-ball paradigm is one of the popular contenders to resolve
the black hole information-loss paradox. The idea here is that an
astrophysical black hole can have a radius few Planck length greater
than then the gravitational radius, i.e. $r_{bh}=r_{g}(1+\epsilon)$,
where $\epsilon<0.5r_{g}=3Gm$ to avoid having an event horizon. There
are already astrophysical constraints on $\epsilon$, see~\citep{Guo:2017jmi,Maggio:2020jml}.}.

The aim of this paper will be two-fold. First of all, we will argue
that gravity is unique in this regard as it provides an infrared scale.
This infrared length scale may even be larger than that gravitational
radius, $L\geq r_{g}=2GM$, we will provide an example of this. A
similar analysis in the quantum electrodynamics (QED) does not yield
any such infrared length scale. In particular, we will show that by
\textit{integrating out an electron} in QED, the photon vacuum is
that of a displaced vacuum (similar to the case of a graviton), but
the occupation number of photon in this case is always bounded below
unity for foreseeable energies. In fact, the occupation number of
photons is proportional to the fine structure constant. Indeed, the
fine structure evolves with the energy in the ultraviolet, but it
remains below unity for energies below the Planck energy. Moreover,
we will further show that Bekenstein's entropy bound in our case is
always satisfied~\citep{Bekenstein}, i.e. Bekenstein's entropy is
always bounded by the energy and the distance scale.

The second goal will be to generalise our earlier results of Ref.~\citep{Bose:2021ekc}
by including both relativistic/non-relativistic effects while integrating
out the energy momentum tensor for the matter field. To illustrate,
we will consider an in-falling thin shell of matter, in an adiabatic
approximation, where the vacuum changes slowly. We will show that
up to the leading order in the Newton's constant, $G$, by integrating
out the energy momentum tensor of an in-falling thin shell, we will
obtain a large occupation number of gravitons. In fact, we will further
show that the infrared scale in this example is \textit{slightly larger
than the Schwarzschild radius} of the corresponding mass $M$, i.e.,
$L\geq r_{g}=2GM$. Moreover, we will argue that surprisingly such
an infrared scaling persists for higher curvature theories of gravity
as well, but now the emergence of a new infrared length scale is \textit{different}
from that of general relativity. This occurs due to the fact that
higher curvature theories of gravity brings in a new mass scale, $M_{s}\leq M_{p}$.

In section II, we will show how by integrating out the electron, we
will obtain that the photon vacuum to be displaced and the corresponding
occupation number would scales as that of the fine structure constant.
In section III, we will generalise our earlier results of Ref.~\citep{Bose:2021ekc}
for the relativistic energy momentum tensor for an arbitrary geometric
configuration, and find the corresponding number of gravitons. In
section IV, we consider an example of an in falling thin shell of
matter and compute the graviton occupation number by integrating out
the thin shell of matter. In section V, we will discuss the infrared
length scale in theories of gravity within general relativity and
in higher curvature theories of gravity.


\section{Number of photons and the fine structure}

Let us now consider the example within QED. Working in a Coulomb gauge
we will expand the electromagnetic field as: 
\begin{equation}
{A}^{i}(\bm{x})=\int\frac{d\bm{k}}{(2\pi)^{3}}\frac{1}{\sqrt{2\omega_{k}}}{a}_{\bm{k},\lambda}\mathtt{e}_{\lambda}^{i}(\bm{n})e^{i\bm{k}\cdot\bm{x}}+\text{H.c.},\label{eq:emfield}
\end{equation}
where $i=1,2,3$, $\bm{x}=(x,y,z)$, $\omega_{k}=k$, $k=\|\bm{k}\|$,
$\bm{n}=\bm{k}/\|\bm{k}\|$, ${a}_{\bm{k},\lambda}$ is the annihilation
operator, and $\mathtt{e}^i_{\lambda}$ denote the basis vectors for
the two polarisations, $\lambda=1,2$. The completeness relation is
given by: 
\begin{equation}
P^{ij}(\bm{n})\equiv\sum_{\lambda}\mathtt{e}_{\lambda}^{i}(\bm{n})\mathtt{e}_{\lambda}^{j}(\bm{n})=\delta^{ij}-\bm{n}^{i}\bm{n}^{j}.\label{eq:completeness}
\end{equation}
The interaction Hamiltonian is given by 
\begin{equation}
H_{\text{int}}=\int d\bm{x}A^{i}(\bm{x})J_{i}(\bm{x}),\label{eq:interaction-1}
\end{equation}
where $J_{i}$ are the components of the current density. We will
now proceed by taking the mean-field approximation
\begin{equation}
J_{i}(\bm{x})\rightarrow\langle J_{i}(\bm{x})\rangle,
\end{equation}
where $\langle J_{i}(\bm{x})\rangle=\text{tr}[\rho J_{i}(\bm{x})]$
is the expectation value of the current density, and $\rho$ is a
generic matter state \footnote{We summarize here the general procedure we will follow for computing
the number of photons/gravitons. We first take the mean-field approximation
of the electromagnetic/gravitational interaction Hamiltonian $H_{\text{int}}$
by \emph{tracing out} the matter state $H_{\text{int}}\rightarrow\langle H_{\text{int}}\rangle$,
where$\langle\,\cdot\,\rangle=\text{tr}[\rho\,\cdot\,]$ indicates
the trace with respect to the matter state $\rho$. We find that the
ground state of the electromagnetic/gravitational field becomes displaced
depending on the values of $\langle J_{i}(\bm{x})\rangle$ and $\langle T_{ij}(\bm{x})\rangle$,
where $J_{i}(\bm{x})$ and $T_{ij}(\bm{x})$ denote the current density
and the stress-energy tensor, respectively. In the computation we
do not specify directly the matter state $\rho$, but only make generic
symmetry and dimensional consideration about the expectation values
$\langle J_{i}(\bm{x})\rangle$ and $\langle T_{ij}(\bm{x})\rangle$.
Assuming such ground states for the electromagnetic/gravitational
field (i.e. displaced coherent states) we then estimate the corresponding
number of photons/gravitons, $N_{p}$ and $N_{g}$, respectively.
\label{fn:summary}}. From Eq.~\eqref{eq:emfield} and \eqref{eq:interaction-1}, we
find 
\begin{equation}
H_{\text{int}}=\int\frac{d\bm{k}}{(2\pi)^{3}}\frac{1}{\sqrt{2\omega_{k}}}{a}_{\bm{k},\lambda}\mathtt{e}_{\lambda}^{i}(\bm{n})\tilde{J}_{i}(\bm{k})+\text{H.c.},\label{eq:derivation1-1}
\end{equation}
where we have introduced the Fourier transform 
\begin{equation}
\tilde{J}_{i}(\bm{k})=\int d\boldsymbol{x}e^{i\bm{k}\cdot\bm{x}}\langle J_{i}(\bm{x})\rangle.
\end{equation}
We will work in the basis where $\mathtt{e}_{\lambda}^{i}(\bm{n})$
is real-valued, but $\tilde{J}_{i}(\bm{k})$ can in general be a c-number.
However, we can always absorb any global phase from the $\sum_{i}\mathtt{e}_{\lambda}^{i}(\bm{n})\tilde{J}_{i}(\bm{k})=\vert C(\bm{k})\vert e^{i\theta(\bm{k})}$
by a redefinition of the modes, i.e. ${a}_{\bm{k},\lambda}e^{i\theta(\bm{k})}\rightarrow{a}_{\bm{k},\lambda}$
(and which leaves invariant kinetic term for a photon in a Coulomb
gauge; $\sim a_{\bm{k},\lambda}^{\dagger}a_{\bm{k},\lambda}$) 
\begin{eqnarray}
H_{p} & = & \int d\boldsymbol{k}\omega_{k}a_{{\bm{k}},\lambda}^{\dagger}a_{{\bm{k},\lambda}}\nonumber \\
 & = & \sum_{\lambda}\int d\boldsymbol{k}\frac{\omega_{k}}{4}\left[P_{\bm{k},\lambda}^{2}+Y_{\bm{k},\lambda}^{2}\right]\,,\label{photon}
\end{eqnarray}
where 
\begin{eqnarray}
Y_{\bm{k},\lambda}=a_{\bm{k},\lambda}^{\dagger}+a_{\bm{k},\lambda}\,,\\
P_{\bm{k},\lambda}=i(a_{\bm{k},\lambda}^{\dagger}-a_{\bm{k},\lambda})\,,
\end{eqnarray}
follow the commutation relation $[a_{\bm{k},\lambda},a_{\bm{k'}\lambda'}^{\dagger}]=\delta({\bm{k}}-{\bm{k}'})\delta_{\lambda\lambda'}$.
From Eq.~\eqref{eq:derivation1-1}, we find 
\begin{equation}
H_{\text{int}}=\int\frac{d\bm{k}}{(2\pi)^{3}}\frac{1}{\sqrt{2\omega_{k}}}\vert\mathtt{e}_{\lambda}^{i}(\bm{n})\tilde{J}_{i}(\bm{k})\vert({a_{\bm{k},\lambda}^{\dagger}+a_{\bm{k},\lambda}}).\label{eq:derivation2-1}
\end{equation}
Specifically, combining the interaction term with the kinetic term
of the electromagnetic field, Eqs.~(\ref{photon}) and (\ref{eq:derivation2-1}),
we find: 
\begin{equation}
{H}_{tot}=\int\frac{d\bm{k}}{(2\pi)^{3}}\frac{\omega_{k}}{4}\Biggl[{P}_{\bm{k},\lambda}^{2}+({Y}_{\bm{k},\lambda}-\alpha_{\bm{k},\lambda})^{2}\Biggr],\label{eq:total2-1}
\end{equation}
where now 
\begin{equation}
\alpha_{\bm{k},\lambda}\equiv\sqrt{\frac{2}{\omega_{k}^{3}}}\vert\mathtt{e}_{\lambda}^{i}(\bm{n})\tilde{J}_{i}(\bm{k})\vert.
\end{equation}
Note that the electromagnetic field $a_{\bm{k},\lambda}$ is in a
ground state, centred around $\alpha_{\bm{k},\lambda}$, which is
described by a \textit{displaced coherent state} of a photon~\citep{Quantum-optics}:
\begin{equation}
|\alpha_{\bm{k},\lambda}\rangle=D(\alpha_{\bm{k},\lambda})|0\rangle=e^{\alpha_{\bm{k},\lambda}\left[a_{\bm{k},\lambda}^{\dagger}+a_{\bm{k},\lambda}\right]}|0\rangle
\end{equation}
We are assuming that the electromagnetic field is in the ground state
of the displaced harmonic trap, and the vacuum is \textit{stable}
and obeys \textit{adiabatic} conditions. Indeed, a different choice
of the vacuum for $A_{i}({\bm{x}})$ will not change significantly
the final result as long as the state remains centred and confined
around the same minimum and obeys adiabaticity, given by $|\alpha_{{\bm{k}},\lambda}\rangle$.

For such a displaced quantum state we can compute the expectation
values $\langle\cdot\rangle$. We will now compute the number operator
$N_{\bm{k},\lambda}=a_{\bm{k},\lambda}^{\dagger}a_{\bm{k},\lambda}$,
where $\langle N_{\bm{k},\lambda}\rangle=|\alpha_{{\bm{k}},\lambda}|^{2}$,
and the total number of photons is summation of all ${\bm{k}},\lambda$
modes : 
\begin{equation}
N_{p}\equiv\sum_{\lambda}\int d\bm{k}\vert\alpha_{\bm{k},\lambda}\vert^{2}=\sum_{\lambda}\int d\bm{k}\frac{\vert\mathtt{e}_{\lambda}^{i}(\bm{n})\tilde{J}_{i}(\bm{k})\vert^{2}}{4\pi^{3}\omega_{k}^{3}}.\label{eq:d1-1}
\end{equation}
Let us then perform the sum over the polarizations. Exploiting the
completeness relation in Eq.~\eqref{eq:completeness}, we find 
\begin{alignat}{1}
\sum_{\lambda}\vert\mathtt{e}_{\lambda}^{i}(\bm{n})\tilde{J}_{i}(\bm{k})\vert^{2} & =\sum_{i,j,\lambda}\mathtt{e}_{\lambda}^{i}(\bm{n})\mathtt{e}_{\lambda}^{j}(\bm{n})\tilde{J}_{i}(\bm{k})[\tilde{J}_{j}(\bm{k})]^{*}\nonumber \\
 & =\sum_{i,j}P^{ij}(\bm{n})\tilde{J}_{i}(\bm{k})[\tilde{J}_{j}(\bm{k})]^{*}.\label{eq:d2}
\end{alignat}
By inserting Eq.~\eqref{eq:d2} back in Eq.~\eqref{eq:d1-1}, we
finally find: 
\begin{equation}
N_{p}=\int d\bm{k}\frac{1}{4\pi^{3}\omega_{k}^{3}}P^{ij}(\bm{n})\tilde{J}_{i}(\bm{k})[\tilde{J}_{j}(\bm{k})]^{*}.\label{eq:d4-1}
\end{equation}
At this point of the calculation we have not made any assumptions
on the current density and thus Eq.~\eqref{eq:d4-1} is still completely
general.

Let us now make some simplifying assumptions. We assume that the Fourier
transform of the current can be split into the radial and angular
components 
\begin{equation}
\tilde{J}_{i}(\bm{k})=R_{i}(\omega_{k})\Omega_{i}(\bm{n}),\label{eq:12}
\end{equation}
where $\omega_{k}=k$. We then insert back Eq.~\eqref{eq:12} into
Eq.~\eqref{eq:d4-1}, and use $d\bm{k}=dkd\bm{n}=\frac{\omega_{k}^{2}}{c}d\omega_{k}d\bm{n}$,
to find 
\begin{equation}
N_{p}=\int d\omega_{k}\frac{R_{i}(\omega_{k})[R_{j}(\omega_{k})]^{*}}{4\pi^{3}\omega_{k}}\int d\bm{n}P^{ij}(\bm{n})\Omega_{i}(\bm{n})[\Omega_{j}(\bm{n})]^{*}.\label{eq:d4-11}
\end{equation}
Without loss of generality, let us furthermore assume that we get
a nonzero contribution only for $i=j=3$, and $R(\omega_{k})\equiv R_{i}(\omega_{k})$.
The number of photons after tracing out an electron simplifies to
\begin{equation}
N_{p}=\int d\omega_{k}\frac{\vert R(\omega_{k})\vert^{2}}{4\pi^{3}\omega_{k}}\underbrace{\int d\bm{n}P^{ii}(\bm{n})\vert\Omega_{i}(\theta,\phi)\vert^{2}}_{\sim\mathcal{O}(1)}.\label{eq:d4-1-1}
\end{equation}
By assuming that the angular part will be non-vanishing, we are thus
left with 
\begin{equation}
N_{p}\sim\int d\omega_{k}\frac{\vert R(\omega_{k})\vert^{2}}{4\pi^{3}\omega_{k}}\,.\label{eq:Np}
\end{equation}
We now make the assumption that 
\begin{equation}
R(\omega_{k})\sim eL\omega_{k},
\end{equation}
where $L$ is the side of the box containing the current with total
charge $e$ and introduce the frequency cutoff 
\begin{equation}
\bar{\omega}=\frac{2\pi}{L}.
\end{equation}
Here we have in mind the following: $L\omega_{k}$ is a velocity and
thus $R(\omega_{k})\sim eL\omega_{k}$ is a current. We can consider
other examples too, for example, for an in falling charged sphere
we will have 
\begin{equation}
R(\omega_{k})\sim i\frac{eL\omega_{k}}{3}\label{eq:23}
\end{equation}
and $\Omega_{3}(\theta,\phi)=1$, which gives $\int d\bm{n}P_{33}(\bm{n})\vert\Omega_{3}(\theta,\phi)\vert^{2}=\frac{8\pi}{3}$. 

In the above equations the length $L$ was introduced from dimensional
analysis. However, it can be shown that such a length-scale $L$ also
arises in specific examples. For example, one can consider a thin
charged shell of radius $R_{0}$ and compute the Fourier transform
of the currents $\tilde{J}_{i}(\bm{k})$ (Appendix A). Moreover, by
expanding the currents to order $\mathcal{O}(\omega_{k})$ we find
that the only non-vanishing component is given by $\tilde{J}_{3}(k)\approx i\frac{e\omega_{k}R_{0}}{3}$.
From Eqs.~\eqref{eq:12} and \eqref{eq:23} we can then conclude
that $L$ can be identified with $R_{0}$ in such an example.

Now, if we combine all these results, we find from Eq.~\eqref{eq:Np}:
\begin{equation}
N_{p}\sim\frac{e^{2}L^{2}}{4\pi^{3}}\int_{0}^{\bar{\omega}}d\omega_{k}\omega_{k}.\label{eq:Np-1}
\end{equation}
We thus finally find: 
\begin{equation}
N_{p}\sim\frac{e^{2}}{4\pi}=\alpha_{em}\ll1.\label{eq:alpha}
\end{equation}
We have obtained an interesting result; by tracing out the current
driven by one electron, the photon occupation number is \textit{just}
that of the fine structure constant (see also \citep{Muck} for a
computation of the number of longitudinal photons). The QED interactions
will allow the fine structure constant to evolve with energies. Up
to all relevant energies, i.e., say up to the Planck scale, the fine
structure constant remains $\alpha_{em}(\mu)\ll1$, where $\mu$ is
the momentum scale. 

We can gain further insight into this problem by recalling that the
Bekenstin's entropy~\citep{Bekenstein} for any system is always
bounded by the energy and the distance scale, i.e. $S_{BEK}\leq2\pi ER$.
We can check this in our case; say for instance, we can estimate the
entropy for a thin shell. The energy of a charge carrying thin shell
of radius $R$ is given by, $E\sim\frac{e^{2}}{8\pi R}$. Therefore,
the entropy is given by: 
\begin{equation}
S_{BEK}\sim e^{2}\sim\alpha_{em}\sim N_{p}.
\end{equation}
Bekenstein's entropy is now related to the number of photons obtained
by tracing out the charge. What it also suggests that the electron
is inherently a quantum system, it can never be classicalised for
a foreseeable energy, e.g. the Planck scale~\citep{Gockeler:1997dn}.


\section{Tracing out energy momentum tensor -- relativistic treatment}

In this section, we will study the graviton occupation number $N_{g}$,
in a long wavelength limit, where we can perturb the metric by: 
\begin{equation}
g_{\mu\nu}=\eta_{\mu\nu}+h_{\mu\nu},\label{eq:expansion}
\end{equation}
where $\mu,\nu=0,1,2,3$. Note that here we have perturbed the metric
around Minkowski background, but we will mention below other choices
of the background.

We will start the gravitational field in the transverse traceless
(TT) gauge in the asymptotically flat region of space time: 
\begin{equation}
{h}^{ij}(\bm{x})=\sum_{\lambda}\int d\bm{k}\sqrt{\frac{G}{\pi^{2}\omega_{k}}}{g}_{\bm{k},\lambda}\mathtt{e}_{\lambda}^{ij}(\bm{n})e^{i\bm{k}\cdot\bm{x}}+\text{H.c.},\label{eq:gravfield}
\end{equation}
where $i,j=1,2,3$, $G$ is the Newton's constant, $\omega_{k}=k$,
$k=\|\bm{k}\|$, $\bm{n}=\bm{k}/\|\bm{k}\|$, ${g}_{\bm{k},\lambda}$
is the annihilation operator, and $\mathtt{e}_{\lambda}^{ij}$ denote
the basis tensors for the two polarisations, $\lambda=1,2$. We can
write the Hamiltonian governed by the kinetic term of the massless
graviton field to be: 
\begin{eqnarray}
H_{\text{grav}} & = & \int d\bm{k}\,\omega_{k}g_{\bm{k},\lambda}^{\dagger}{g}_{\bm{k},\lambda}\nonumber \\
 & = & \sum_{\lambda}\int d\bm{k}\frac{\omega_{k}}{4}\left[{P}_{\bm{k},\lambda}^{2}+{Y}_{\bm{k},\lambda}^{2}\right],\label{eq:Hgrav}
\end{eqnarray}
where 
\begin{equation}
{Y}_{\bm{k},\lambda}={g}_{\bm{k},\lambda}+{g}_{\bm{k},\lambda}^{\dagger}\qquad{P}_{\bm{k},\lambda}=i({g}_{\bm{k},\lambda}^{\dagger}-{g}_{\bm{k},\lambda}).\label{eq:2}
\end{equation}
They operators follow the commutation relation $[g_{\bm{k},\lambda},g_{\bm{k'}\lambda'}^{\dagger}]=\delta({\bm{k}}-\bm{k'})\delta_{\lambda\lambda'}$.
The minimal coupling between graviton and matter dictates the interaction
Hamiltonian $\int\sqrt{-g}d^{4}xG^{\mu\nu}T_{\mu\nu}$, and can be
written in the TT-gauge as: 
\begin{equation}
H_{\text{int}}=-\frac{1}{2}\int d\bm{x}h^{ij}(\bm{x})T_{ij}(\bm{x}),\label{eq:interaction}
\end{equation}
where $\bm{x}=(x,y,z)$, and $T_{ij}$ are the components of the stress-energy
tensor. We will now take the mean-field approximation, where we take
\begin{equation}
T_{ij}(\bm{x})\rightarrow\langle T_{ij}(\bm{x})\rangle,
\end{equation}
 where $\langle T_{ij}(\bm{x})\rangle=\text{tr}[\rho T_{ij}(\bm{x})]$
is the expectation value of the stress-energy tensor, and $\rho$
is a generic matter state \footref{fn:summary}. 

From Eq.~\eqref{eq:gravfield} and \eqref{eq:interaction}, we find
\begin{equation}
H_{\text{int}}=-\frac{1}{2}\sum_{\lambda}\int d\bm{k}\sqrt{\frac{G}{\pi^{2}\omega_{k}}}{g}_{\bm{k},\lambda}\mathtt{e}_{\lambda}^{ij}(\bm{n})\tilde{T}_{ij}(\bm{k})+\text{H.c.},\label{eq:derivation1}
\end{equation}
where we have introduced the Fourier transform 
\begin{equation}
\tilde{T}_{ij}(\bm{k})=\int d\bm{x}\langle T_{ij}(\bm{x})\rangle e^{i\bm{k}\cdot\bm{x}}.
\end{equation}
We will work in the basis where $\mathtt{e}_{\lambda}^{ij}(\bm{n})$
is a real-valued, but $\tilde{T}_{ij}(\bm{k})$ in general can be
a c-number. We can always absorb any global phase from the $\sum_{ij}\mathtt{e}_{\lambda}^{ij}(\bm{n})\tilde{T}_{ij}(\bm{k})=\vert A(\bm{k})\vert e^{i\theta(\bm{k})}$
by a redefinition of the modes, i.e. ${g}_{\bm{k},\lambda}e^{i\theta(\bm{k})}\rightarrow{g}_{\bm{k},\lambda}$
(and which leaves invariant kinetic term $\sim g_{\bm{k},\lambda}^{\dagger}g_{\bm{k},\lambda}$).
From Eq.~\eqref{eq:derivation1}, we thus find 
\begin{equation}
H_{\text{int}}=-\frac{1}{2}\sum_{\lambda}\int d\bm{k}\sqrt{\frac{G}{\pi^{2}\omega_{k}}}\vert\mathtt{e}_{\lambda}^{ij}(\bm{n})\tilde{T}_{ij}(\bm{k})\vert{(g_{\bm{k},\lambda}+g_{\bm{k},\lambda}^{\dagger})},\label{eq:derivation2}
\end{equation}
By combining the interaction term with the kinetic term of the gravitational
field, we obtain: 
\begin{equation}
{H}=\sum_{\lambda}\int d\bm{k}\frac{\omega_{k}}{4}\Biggl[{P}_{\bm{k},\lambda}^{2}+({Y}_{\bm{k},\lambda}-\alpha_{\bm{k},\lambda})^{2}\Biggr],\label{eq:total2}
\end{equation}
where now 
\begin{equation}
\alpha_{\bm{k},\lambda}\equiv\sqrt{\frac{G}{\pi^{2}\omega_{k}^{3}}}\vert\mathtt{e}_{\lambda}^{ij}(\bm{n})\tilde{T}_{ij}(\bm{k})\vert.
\end{equation}
Assuming that the the gravitational field is in the ground state of
Eq.~\eqref{eq:total2}, i.e. in a displaced coherent state, similar
to the electromagnetic case~\citep{Quantum-optics}: 
\begin{equation}
|\alpha_{\bm{k},\lambda}\rangle=D(\alpha_{\bm{k},\lambda})|0\rangle=e^{\alpha_{\bm{k},\lambda}\left[g_{\bm{k},\lambda}^{\dagger}+g_{\bm{k},\lambda}\right]}|0\rangle\,,
\end{equation}
we can compute the number operator $N_{\bm{k},\lambda}=g_{\bm{k},\lambda}^{\dagger}g_{\bm{k},\lambda}$,
where $\langle N_{\bm{k},\lambda}\rangle=|\alpha_{{\bm{k}},\lambda}|^{2}$,
and the total number of gravitons by summing over all ${\bm{k}},\lambda$
modes: 
\begin{equation}
N_{g}\equiv\sum_{\lambda}\int d\bm{k}\vert\alpha_{\bm{k},\lambda}\vert^{2}=\sum_{\lambda}\int d\bm{k}\frac{G}{\pi^{2}\omega_{k}^{3}}\vert\mathtt{e}_{\lambda}^{ij}(\bm{n})\tilde{T}_{ij}(\bm{k})\vert^{2},\label{eq:d1}
\end{equation}
where we sum over the polarizations and momenta of each mode. Let
us first perform the sum over the polarizations~\footnote{We recall the basis tensors satisfy the completeness relation: $P^{ijkl}\equiv\sum_{\lambda}\mathtt{e}_{\lambda}^{ij}(\bm{n})\mathtt{e}_{\lambda}^{kl}(\bm{n})=P^{ik}P^{jl}+P^{il}P^{jk}-P^{ij}P^{kl}$
where $P^{ij}\equiv P^{ij}(\bm{n})=\delta^{ij}-\bm{n}^{i}\bm{n}^{j}$.} by exploiting the completeness relation, we find 
\begin{equation}
N_{g}=\int d\bm{k}\frac{G}{\pi^{2}\omega_{k}^{3}}P^{iji'j'}(\bm{n})\tilde{T}_{ij}(\bm{k})[\tilde{T}_{i'j'}(\bm{k})]^{*}.\label{eq:d4}
\end{equation}
This is the generalisation of our earlier computation~\citep{Bose:2021ekc},
where we have not made any assumptions on the stress-energy tensor
and thus Eq.~\eqref{eq:d4} is applicable for any matter, i.e. a
relativistic or a non-relativistic equation of state.


\section{In-falling shell}

Since we have the most general expression, let us consider a special
example of a radially in-falling thin shell of matter, whose stress-energy
tensor is given by, see~\citep{Poisson:2009pwt}: 
\begin{alignat}{1}
T_{ij} & (\bm{R})\equiv{\varepsilon\delta(R-R_{0})}n_{i}(\theta,\phi)n_{j}(\theta,\phi),
\label{eq:Tij}
\end{alignat}
where $\varepsilon$ is a surface energy density, $\bm{R}=(R,\theta,\phi)$
is the 3-vector expressed in spherical coordinates, and $\bm{n}=(\text{sin}(\theta)\text{cos}(\phi),\text{sin}(\theta)\text{sin}(\phi),\text{cos}(\theta))$
is the unit vector. The radius of the thin shell is $R_{0}\equiv R_{0}(t)$
with $\dot{R}_{0}<0$ ($\dot{R}_{0}>0$) corresponding to an in falling
(outgoing) shell), where dot denotes time derivative with respect
to time $t$. The surface energy density, here assumed homogeneous
for simplicity, can be written in terms of an effective mass $M$:
\begin{equation}
\varepsilon=\frac{M}{4\pi R_{0}(t)^{2}}.\label{eq:surface_energy}
\end{equation}
Inside the shell the gravitational potential is zero, while outside
the shell the gravitational metric potential behaves as a Schwarzschild
metric if $R_{0}$ is fixed, if not, it should be similar to the in
falling shell of a Vaidya metric~\citep{Poisson:2009pwt}.

To simplify our computations, and just to capture the leading order
effect in $G$, we will work in the regime where the gravitational
potential is negligible, i.e. $R_{0}\gg r_{g}=2GM$. The 3D Fourier
transform of the stress-energy tensor for the static shell is defined
as: 
\begin{alignat}{1}
\tilde{T}_{ij}(\bm{k}) & \equiv\int d\bm{R}\,T_{ij}(\bm{R})e^{i\bm{k}\cdot\bm{R}}.\label{eq:FourierTransform}
\end{alignat}
We can evaluate such integrations using spherical coordinates and
by choosing to align the $\bm{k}$ vector with the $z$ axis (Appendix
B). Specifically, inserting Eq.~\eqref{eq:Tij} in Eq.~\eqref{eq:FourierTransform}
we then find after expanding up to order $\mathcal{O}(k^{4})$, 
\begin{alignat}{1}
\tilde{T}_{11}(k) & =\tilde{T}_{22}(k)=M(\frac{1}{3}-\frac{R_{0}^{2}k^{2}}{30}+...),\\
\tilde{T}_{33}(k) & =M(\frac{1}{3}-\frac{R_{0}^{2}k^{2}}{10}+...).\label{eq:t11t22t33}
\end{alignat}
Here we will be interested in the regime $R_{0}(t)k\ll1$ where higher
order terms can be neglected. Using $k=\omega_{k}$ this corresponds
to the UV cut-off frequency, $\omega_{k}\ll1/R_{0}(t)$. Note that
although $R_{0}(t)$ has an explicit time dependence, here we will
only consider that the in-falling shell is moving very slowly and
the corresponding vacuum for the shell remains adiabatic. Violation
of adiabaticity in the vacuum will lead to particle creation and the
break down of our key assumptions.


This interaction Hamiltonian considered here is consistent with the
perturbations around Minkowski spacetime $\eta_{\mu\nu}$ in Eq.~\eqref{eq:expansion}.
As we will see below, this will lead to $N_{g}\sim\mathcal{O}(h_{\mu\nu}^{2})\sim\mathcal{O}(G)$.
In the case of an in-falling shell, the metric is not Minkowski even
for $\omega_{k}\ll1/R_{0},R_{0}k\ll1$. In the static case, the metric
outside the shell is that of the Schwarzschild (for static) or Vaidya
(dynamic) metric. Therefore, the correct expansion of the perturbations
should be: $g_{\mu\nu}=\tilde{g}_{\mu\nu}+\tilde{h}_{\mu\nu}$, where
$\tilde{g}_{\mu\nu}$ is the background metric.

However, in the linearised limit, the leading order term in the metric
remains that of Minkowski one for $r\sim R_{0}\gg r_{g}=2GM$, so
any correction due to $2GM/R$ contribution in the metric will yield
higher order corrections in $G$, i.e. due to $\sqrt{-g}$ contribution
in the interaction Hamiltonian, $\int\sqrt{-g}d^{4}xG^{\mu\nu}T_{\mu\nu}$.
Hence, the above mentioned corrections due to either Schwarzschild
or Vaidya remain sub-leading in $N_{g}\sim{\cal O}(G^{2})$.

Therefore, working at the leading order in $G$, we can split Eq.~\eqref{eq:d4}
into the radial and angular parts: 
\begin{equation}
N_{g}=\frac{G}{\pi^{2}}\int\frac{d\omega_{k}}{\omega_{k}}\vert\tilde{T}(\omega_{k})\vert^{2}\int d\bm{n}P^{iji'j'}(\bm{n})F_{ij}(\bm{n})F_{i'j'}(\bm{n}),\label{eq:d44}
\end{equation}
where we have used $d\bm{k}=dkd\bm{n}=\omega_{k}^{2}d\omega_{k}d\bm{n}$,
and $\tilde{T}(\omega_{k})$ and $F_{ij}(\bm{n})$ will be specified
below. The expansion of the stress energy tensor from Eq.~\eqref{eq:t11t22t33}
can be treated perturbatively -- computing first the contribution
to order $\mathcal{O}(1)$, and then to order $\mathcal{O}(k^{2})$.

From Eq.~\eqref{eq:t11t22t33}, at the lowest order $\mathcal{O}(1)$,
we have the terms 
\begin{alignat}{1}
\tilde{T}(\omega_{k}) & =\frac{M}{3},\qquad F_{ij}=\delta_{ij},
\end{alignat}
which however leads to a vanishing contribution in Eq.~\eqref{eq:d4}
as we have $\sum_{i,j}\int d\bm{n}P^{iijj}(\bm{n})=0$ due to symmetry
considerations. From Eq.~\eqref{eq:t11t22t33}, the quadratic terms,
$\mathcal{O}(k^{2})$, give 
\begin{alignat}{1}
\tilde{T}(\omega_{k}) & =\frac{MR_{0}^{2}\omega_{k}^{2}}{30},
\end{alignat}
which can be seen as the relativistic counterpart of a harmonic oscillator
potential energy. In addition, the only nonzero angular terms are
$F_{11}=1,$$F_{22}=1$, but $F_{33}=3$ (the asymmetry of $F_{33}$
with respect to $F_{11},F_{22}$ originates from the choice of the
alignment of the $\bm{k}$ vector with the z axis). After some straightforward
but tedious algebra we eventually find 
\begin{equation}
N_{g}=\frac{G}{\pi^{2}c^{5}}\int\frac{d\omega_{k}}{\omega_{k}}\vert{\tilde{T}(\omega_{k})}\vert^{2}{\int d\bm{n}P^{iji'j'}(\bm{n})F_{ij}(\bm{n})F_{i'j'}(\bm{n})},\label{eq:Ng2}
\end{equation}
We now introduce an UV cutoff in Fourier space, i.e. $\bar{\omega}=\frac{2\pi}{R_{0}}$
(matching the approximations in Eq.~\eqref{eq:t11t22t33}). From
Eq.~\eqref{eq:Ng2}, we then readily find 
\begin{equation}
N_{g}=\frac{32GR_{0}^{4}M^{2}}{3375\pi}\int_{0}^{\bar{\omega}}d\omega_{k}\omega_{k}^{3}.\label{eq:d12}
\end{equation}
Finally, by evaluating the frequency integral, and using $\bar{\omega}=\frac{2\pi}{R}$,
we obtain the occupation number at the leading order in $G$ to be:
\begin{equation}
N_{g}=\frac{128\pi^{3}}{3375}GM^{2},\label{eq:d13}
\end{equation}
which is very similar to what we had obtained earlier in the non-relativistic
setup~\citep{Bose:2021ekc}. The numerical factors are indeed different,
due to the geometry, but $GM^{2}$ factor remains the same.

Recalling our discussions in section \ref{sec:-Infrared-scale}, we
can establish a new length scale in the infrared. Since, $l_{P}=\sqrt{G}$,
then with the help of Eq.~\eqref{eq:d13} and Eq.(\ref{P-0}), we
obtain: 
\begin{equation}
L\leq\sqrt{N_{g}}l_{P}=\sqrt{\frac{128\times16\times\pi^{4}}{3375}}GM\sim3.8r_{g},\label{eq:d13-2}
\end{equation}
which signifies that the quantum effects, such as the quantum fluctuations
of the virtual gravitons play an interesting role as the shell crosses
$\sim3.8r_{g}$. This is a significant result, which matches the expectations
found in the analysis of the fuzz-ball scenario where the fuzz-ball
micro states played an important role even before the black hole horizon
started forming~\citep{Mathur:2020ely}. Note that we have not computed
the amplitude of the total probability here in Eq.~(\ref{P-0}).
However, the exponent being more sensitive, we expect that the emergence
of a new scale in gravity is inevitable whenever there exists a large
number of states, i.e. $N_{g}\gg1$. The actual numerical factor $3.8r_{g}$
may change, but within an order of magnitude of order $G$ our conclusions
will remain intact. Appearance of a new scale in the infrared is a
welcoming sign, in particular, it enforces us to rethink our understanding
of a traditional classical black hole with an event horizon~\citep{Mathur:2005zp}.
Given the future advancements in observational gravitational waves,
our analysis prompts us to study various consequences of a compact
object devoid of any classical horizon, see~\citep{Guo:2017jmi,Maggio:2020jml,Koshelev:2017bxd}.

For the rest of the paper, we will consider the physical effects due
to higher curvature contributions, such as ${\cal R}^{2}$ in the
gravitational action Eq.~(\ref{eq:Sq}).


\section{Infrared scale for gravitons\label{sec:-Infrared-scale}}

\label{NL-0}

Let us begin by considering the gravitational action which also allows
higher order derivative and curvature terms. 
\begin{eqnarray}
S & = & S_{EH}+S_{q}\,,\\
S_{EH} & = & \frac{1}{16\pi G}\int d^{4}x\sqrt{-g}R\,,\\
S_{q} & = & \frac{1}{16\pi G}\int d^{4}x\sqrt{-g}\left[\alpha{\cal R}^{2}+\beta{\cal R}^{3}+\cdots\right]\,,\label{eq:Sq}
\end{eqnarray}
where $M_{p}^{2}=1/(16\pi G)$, $\alpha,\beta$ are dimension full
constants. Since gravity is a massless theory; therefore space time
diffeomorphism invariance allows higher curvature and higher derivative
(more than two covariant derivatives) contributions. In general, we
may expect~\citep{Biswas:2011ar}: 
\begin{equation}
{\cal R}^{2}=Rf(\Box_{s})R+\cdots,
\end{equation}
where $\Box_{s}=\Box/M_{s}^{2}$ operator will bring a new scale $M_{s}$,
where $M_{s}\leq M_{p}$. For all practical purposes we can take $M_{s}=M_{p}$.
We will show that despite of all these scales, there exists a unique
scale in the infrared~\footnote{Note that the graviton propagator depends on the action and the background,
see~\citep{Biswas:2011ar,Biswas:2013kla,DHoker:1999bve}. Higher
derivative contributions to gravity also bring ghosts in the graviton
propagator~\citep{dof,Biswas:2013kla}. In order to avoid ghosts,
either we restrict our action to only two derivatives or all infinite
covariant derivatives, as shown in papers~\citep{Biswas:2011ar,Modesto}.
Such class of gravitational theories are known as infinite derivative
theory of gravity (IDG). Infinite derivatives do not have a point
support, see~\citep{Buoninfante:2018xiw}, and therefore introduces
non-local interactions.} as pointed out in Refs.~\citep{Ruffini,Dvali2013black,Mathur:2005zp,Mathur:2020ely,Bose:2021ekc}.

Let us introduce a characteristic length scale $L$, so that one has
$\partial x\sim L$ and $1/\partial x\sim1/L$; in the same way all
curvature tensors will scale as ${\cal R}\sim1/L^{2},~{\cal R}^{2}\sim1/L^{4},\cdots$.
Therefore, Einstein-Hilbert action will contribute: 
\begin{equation}
S_{EH}\sim M_{p}^{2}L^{2}\,,
\end{equation}
To understand the appearance of this new scale in the infrared, there
is another intuitive way to proceed~\citep{Mathur:2020ely,Mathur-1}
(see also \citep{Zurek:2021} for a discussion motived by the AdS/CFT
calculations).

Let us recall the arguments proposed in Ref.~\citep{Mathur:2020ely,Mathur:2008kg}.
The fluctuations in energy for a given mass is given by; $E\sim M$,
if the fluctuations exist for a time scale $T\sim L$, then the action
will be given by: $S\sim ET$. In \citep{Mathur:2020ely,Mathur:2008kg},
it was argued that for a black hole with a mass $M$ will also accompany
the virtual excitations of the vacuum. Typically, these fluctuations
are \textit{exponentially suppressed}, nevertheless, the number of
states available in the case of a black hole is exponentially large,
as in the case of a fuzz ball. Take only Einstein-Hilbert contribution
in the gravitational action, we will get $S_{EH}\sim ET\sim M_{p}^{2}L^{2}$.
For the black hole case, $L=T=r_{g}=2GM$. Therefore, the total probability
for the existence of a black hole must also take into account of the
gravitational states available, given by ${\cal N}\approx e^{N_{g}}$,
and the gravitational action. Collectively, we can express the total
probability to be: 
\begin{equation}
P_{T}\sim e^{N_{g}}e^{-S}\sim e^{N_{g}-M_{p}^{2}L^{2}}\sim{\cal O}(1)\,.\label{P-0}
\end{equation}
In \citep{Mathur:2020ely,Mathur:2008kg}, the number of states available
for a black hole was due to the fuzz-ball states. In our case, the
number of quantum states available is mainly due to the graviton states,
very similar to~\citep{Ruffini,Dvali2013black,Casadio:2015,Bose:2021ekc}.
The total probability becomes $P_{T}\sim{\cal O}(1)$, provided 
\begin{equation}
L\leq\frac{\sqrt{N_{g}}}{M_{p}}\Longrightarrow M_{eff}\sim\frac{M_{p}}{\sqrt{N_{g}}},\label{P-2}
\end{equation}
where we have taken $L\sim M_{eff}^{-1}$. In this regard fuzz-ball~\citep{Mathur:2005zp}
and the corpuscular hypotheses~\citep{Ruffini,Dvali2013black,Casadio:2015,Bose:2021ekc}
gave very similar results. Both the hypotheses saturate the bound
for $N_{g}$ for the gravitational radius; $L=r_{g}=2GM$. Although,
the fuzz-ball hypothesis will go beyond these steps, and will bring
a new infrared scale which can be even larger than the gravitational
radius. We will discuss them in the next section.

To evaluate $N_{g}$, we will take inspiration from our recent computation
in Ref.~\citep{Bose:2021ekc}, where we have shown that it is indeed
possible to estimate $N_{g}$ by tracing out the non-relativistic
matter, and we had shown that the minimal coupling between the matter
and the gravity would suffice to show that the gravitons are in a
displaced coherent state, whose occupation is given by the same as
that of Bekenstein's entropy bound. 
\begin{equation}
N_{g}=S_{BEK}=\frac{Area}{4G}\sim GM^{2}\sim\left(\frac{M}{M_{p}}\right)^{2}.
\end{equation}
By substituting in Eq.(\ref{P-0}), we obtain the infrared scale in
gravity to be similar to the Schwarzschild radius, i.e. $L\sim r_{g}$,
as we have discussed above. It is worth highlighting that the gravitational
entropy is indeed holographic in nature also, see \citep{thooft,Susskind}.

It is worth comparing out that the electromagnetic case has no infrared
length scale. A similar analysis will suggest that the electromagnetic
action has no explicit length scale dependence, unlike gravity, where
the interaction strength $\sqrt{G}\sim1/M_{p}$, and possess a length
scale. This naturally forbids appearance of an infrared scaling in
the case of a photon.

A natural question arises; could we evaluate $N_{g}$ for other equations
of state, or for other geometries. Our previous computation in Ref.~\citep{Bose:2021ekc}
was performed in a non-relativistic setting. We wish to now consider
these issues and compute $N_{g}$, where we assume that the graviton
vacuum is always \textit{stable} and obeys \textit{adiabatic} condition,
by tracing out the energy momentum tensor of the matter sector. 

Let us consider terms which are proportional to ${\cal R}^{2}$ in
$S_{q}$ in Eq.~(\ref{eq:Sq})~\footnote{The ghost free quadratic curvature action with analytic operators
is given by: $S=1/(16\pi G)\int d^{4}x\sqrt{-g}[R+\alpha(Rf_{1}(\Box_{s})R+R_{\mu\nu}f_{2}(\Box_{s})R^{\mu\nu}+R_{\mu\nu\lambda\sigma}f_{3}(\Box_{s})R^{\mu\nu\lambda\sigma})]$.
The whole action can be made ghost-free in Minkowski~\citep{Biswas:2011ar}
and in maximally symmetric backgrounds. In Minkowski spacetime, the
ghost-free condition demands that $2f_{1}+f_{2}+2f_{3}=0$.}. Our previous arguments from section \ref{sec:-Infrared-scale} have
suggested that on dimensional grounds, the total gravitational action
including Einstein-Hilbert term and the quadratic piece will give
us: 
\begin{equation}
S\sim M_{p}^{2}L^{2}\left[1+\frac{1}{M_{s}^{2}L^{2}}\right]\,,
\end{equation}
If we demand that $S_{q}>S_{EH}$ in Eq.~(\ref{eq:Sq}) for a certain
length scale $L$, then this would mean $L<1/M_{s}$. Now, if we demand
that the virtual excitations of gravitons for a given mass $M$ ought
to play a significant role, then 
\begin{equation}
P_{T}\sim e^{N_{g}}e^{-M_{p}^{2}L^{2}[1+\frac{1}{M_{s}^{2}L^{2}}]}\sim{\cal O}(1)\,,\label{High-P}
\end{equation}
Therefore, if the quadratic in curvature term were to dominate over
Einstein-Hilbert term then the probability would become order one,
provided 
\begin{equation}
M_{s}=\frac{M_{p}}{\sqrt{N_{g}}}\label{IR-0}
\end{equation}
\footnote{This is indeed a very interesting relationship, as we had shown in
a completely different context; how a new scale appears in gravity
but in the context o fa non-local gravitational interaction~\citep{Koshelev:2017bxd}. }To be consistent, we would need $L<M_{s}^{-1}$. Therefore, we will
have the following relationship for an in falling thin shell; 
\begin{equation}
\frac{1}{L}\geq M_{s}=\frac{M_{p}}{\sqrt{N_{g}}}\Longrightarrow L\leq3.8r_{g}\,.
\end{equation}
Let us now check what would happen if we were to demand the domination
of the cubic order terms in the curvature over all the other contributions
for some length scale $L$, then 
\begin{equation}
S\sim M_{p}^{2}L^{2}\left[1+\frac{1}{M_{s}^{2}L^{2}}+\frac{1}{M_{s}^{4}L^{4}}+\cdots\right]\,,
\end{equation}
Indeed, all the higher curvature terms do become important for $L\leq M_{s}^{-1}$,
barring any fine-tuned cancellations. Let's take the cubic term first.
If the cubic contribution dominates overall, then Eq.~(\ref{P-0},
\ref{High-P}) would suggest 
\begin{equation}
L\sim\frac{M_{p}}{M_{s}^{2}}\frac{1}{\sqrt{N_{g}}}\,.\label{IR-HD}
\end{equation}
Indeed, now we have two new parameters to constrain $M_{s}$ and $L$
for a given $N_{g}$. Let us suppose, conservatively, we take $M_{s}\sim M_{p}/\sqrt{N_{g}}$,
then we would obtain the same conclusion that the infrared scale of
gravity becomes $L\leq\sqrt{N_{g}}/M_{p}$, same as that of Eq.~(\ref{IR-0}),
since $L\sim M_{s}^{-1}$. If $M_{s}$ is considered to be the string
scale, then the hierarchy between $M_{s}$ and $M_{p}$ is related
to the gravitational states.

All these results point towards one very crucial fact that irrespective
of any higher-order curvature and/or higher derivative corrections,
there must appear a new scale of gravity in the infrared, which has
a universal feature given by Eq.(\ref{P-2}), i.e. $M_{eff}\sim M_{p}/\sqrt{N_{g}}$,
where $N_{g}$ denotes the number of graviton states associated with
the mass $M$~\citep{Mathur:2020ely}.

\section{Conclusion}

In this paper we have found two results. First, by tracing out the
charged source, i.e. the electron, we have found that the photon vacuum
is displaced. This is analogous to the displaced coherent state of
a photon vacuum with an occupation number of photons, $N_{p}$, which
scales as the fine structure constant. Since the fine structure constant
remains less than one, it implies that the electron remains a quantum
system for energies below the Planck scale. Furthermore, the photon
number is always bounded by the Bekenstein's entropy bound. All the
computations are based on the adiabatic evolution of the charged source
and the photon vacuum.

The second result, we have shown that the gravitational interaction
with the matter is entirely different as expected. By tracing out
the matter, we found that the graviton vacuum is also displaced. Still,
now the occupation number of gravitons is proportional to the Area.
The current result generalises our previous result~\citep{Bose:2021ekc},
where we have generalised the computation for an arbitrary energy
momentum tensor and beyond Einstein-Hilbert action. Motivated by \citep{Mathur:2020ely,Mathur:2008kg},
we have found that by including the large degeneracy provided by the
occupation number of the gravitons in the displaced vacuum, the infrared
scale emerges. This infrared scale can be larger than the gravitational
radius. In fact, in the simple toy model we have studied, i.e. an
in-falling thin shell of matter, the emergence of the infrared length
scales appears to be $L\leq3.8r_{g}$. We have further noticed that
the appearance of this infrared scale in gravity persists even if
we go beyond Einstein-Hilbert action. Apparently, such a new scale
is always determined by the large occupation number of gravitons in
the displaced vacuum, see Eq.~(\ref{IR-HD}).

Our results prompt us to investigate further open questions such as
the new scale of gravity in a generic collapsing geometry, particularly
in the context of cosmology~\citep{Cadoni}, and in the formation
of an ultra compact object. Would the appearance of a new scale alleviate
cosmological Big Bang singularity or resolve black hole singularity?
Would the appearance of a new scale in gravity alter the way we view
traditional black holes? Would there be associated observational signatures
which can be falsifiable by future gravitational wave detectors? All
these questions remain outstanding, indeed they go beyond the scope
of the current paper, and deserves a detailed investigation.

\section*{Acknowledgements}

We would like to thank Samir Mathur for helpful discussions. MT and
SB would like to acknowledge EPSRC grant No.EP/N031105/1, SB the EPSRC
grant EP/S000267/1, and MT funding by the Leverhulme Trust (RPG-2020-197).
AM's research is funded by the Netherlands Organisation for Science
and Research (NWO) grant number 680-91-119.

\appendix

\section{Thin charged shell of matter}

In the case of a charged shell, we start from the current density:
\begin{equation}
J_{i}(\bm{R})=\frac{e}{4\pi R_{0}^{2}}\delta(R-R_{0})n_{i}(\theta,\phi).
\end{equation}
We are interested in the Fourier transform 
\begin{alignat}{1}
\tilde{J}_{i}(\bm{k})= & \frac{e}{4\pi R_{0}^{2}}\int_{0}^{\infty}dR\int_{-1}^{1}d\text{cos}(\theta)\int_{0}^{2\pi}d\phi\,\nonumber \\
 & R^{2}\delta(R-R_{0})n_{i}(\theta,\phi)e^{ikR\text{cos}(\theta)}
\end{alignat}
Performing the integrations we find the following nonzero terms 
\begin{alignat}{1}
\tilde{J}_{3}(k) & =-\frac{ie(kR_{0}\cos(kR_{0})-\sin(k\text{\ensuremath{R_{0}}}))}{k^{2}R_{0}^{2}}
\end{alignat}
where $k\equiv\vert\bm{k}\vert=\omega_{k}$ is the radial component
of the wave vector. 

\section{Thin neutral shell of matter}

We start from the stress-energy tensor: 
\begin{alignat}{1}
\tilde{T}_{ij}(\bm{k})= & \varepsilon\int_{0}^{\infty}dR\int_{-1}^{1}d\text{cos}(\theta)\int_{0}^{2\pi}d\phi\,\nonumber \\
 & R^{2}\delta(R-R_{0})n_{i}(\theta,\phi)n_{j}(\theta,\phi)e^{ikR\text{cos}(\theta)}.
\end{alignat}
Performing the integrations, and inserting Eq.~\eqref{eq:surface_energy},
we find: 
\begin{alignat}{1}
\tilde{T}_{11}(k) & =M\frac{\sin(kR_{0})-kR_{0}\cos(kR_{0})}{k^{3}R_{0}^{3}},\\
\tilde{T}_{22}(k) & =M\frac{\sin(kR_{0})-kR_{0}\cos(kR_{0})}{k^{3}R_{0}^{3}}\\
\tilde{T}_{33}(k) & =M\frac{\left(k^{2}R_{0}^{2}-2\right)\sin(k\text{\ensuremath{R_{0}}})+2k\text{\ensuremath{R_{0}}}\cos(k\text{\ensuremath{R_{0}}})}{k^{3}\text{\ensuremath{R_{0}}}^{3}}
\end{alignat}
where $k\equiv\vert\bm{k}\vert=\omega_{k}$ is the radial component
of the wave vector.


\begin{thebibliography}{99}
\bibitem{Gupta} S. N. Gupta, Phys. Rev. 96 (1954), 1683-1685

\bibitem{dof} P. Van Nieuwenhuizen, ?On ghost-free tensor lagrangians
and linearized gravi- tation, Nucl.Phys. B60 (1973), 478; T.~Biswas,
T.~Koivisto and A.~Mazumdar, ``Nonlocal theories of gravity: the
flat space propagator,'' {[}arXiv:1302.0532 {[}gr-qc{]}{]}.

\bibitem{Baym} G. Baym and T. Ozawa, Proc. Nat. Acad. Sci. 106 (2009),
3035-3040

\bibitem{Dyson} F. J. Dyson, The World on a String?, review of The
Fabric of the Cosmos: Space, Time, and the Texture of Realitiy by
Brian Greene, New York Review of Books, Volume 51, Number 8, May 13,
(2004); T. Rothman and S. Boughn, Found. Phys. 36 (2006), 1801-1825

\bibitem{Ashoorioon:2012kh} A.~Ashoorioon, P.~S.~Bhupal Dev and
A.~Mazumdar, 
Mod. Phys. Lett. A \textbf{29} (2014) no.30, 1450163 doi:10.1142/S0217732314501636
{[}arXiv:1211.4678 {[}hep-th{]}{]}.

\bibitem{Ng} Y. J. Ng and H. van Dam, Found. Phys. 30 (2000), 795-
805 W. A. Christiansen, Y. J. Ng and H. van Dam, Phys. Rev. Lett.
96 (2006), 051301

\bibitem{Amelino} G. Amelino-Camelia, J. R. Ellis, N. E. Mavromatos,
D. V. Nanopoulos and S. Sarkar, Nature 393 (1998), 763- 765; G. Amelino-Camelia,
Living Rev. Rel. 16 (2013), 5

\bibitem{Parikh} M. Parikh, F. Wilczek and G. Zahariade, Int. J.
Mod. Phys. D 29 (2020) no.14, 2042001

\bibitem{Toros:2020dbf} M.~Toro\v{s}, T.~W.~Van De Kamp, R.~J.~Marshman,
M.~S.~Kim, A.~Mazumdar and S.~Bose, 
Phys. Rev. Res. \textbf{3} (2021) no.2, 023178 doi:10.1103/PhysRevResearch.3.023178
{[}arXiv:2007.15029 {[}gr-qc{]}{]}.

\bibitem{Bose:2017nin} S.~Bose, A.~Mazumdar, G.~W.~Morley, H.~Ulbricht,
M.~Toro\v{s}, M.~Paternostro, A.~Geraci, P.~Barker, M.~S.~Kim and
G.~Milburn, 
Phys. Rev. Lett. \textbf{119} (2017) no.24, 240401 doi:10.1103/PhysRevLett.119.240401
{[}arXiv:1707.06050 {[}quant-ph{]}{]}.

\bibitem{Vedral} C. Marletto, V. Vedral, Phys. Rev,Lett. 119, 240402
(2017).

\bibitem{Marshman:2019sne} R.~J.~Marshman, A.~Mazumdar and S.~Bose,
Phys. Rev. A \textbf{101} (2020) no.5, 052110 doi:10.1103/PhysRevA.101.052110
{[}arXiv:1907.01568 {[}quant-ph{]}{]};

\bibitem{Ruffini} R. Ruffini and S. Bonazzola Phys. Rev. 187, 1767
(1969).

\bibitem{Dvali2013black} G. Dvali and C. Gomez, Fortsch. Phys. 61
(2013), 742- 767; G. Dvali and C. Gomez, arXiv:1212.0765 {[}hep- th{]};
G. Dvali, D. Flassig, C. Gomez, A. Pritzel and N. Wintergerst, Phys.
Rev. D 88, no. 12, 124041 (2013), {[}arXiv:1307.3458 {[}hep-th{]}{]};
G. Dvali and C. Gomez, Eur. Phys. J. C 74, 2752 (2014), {[}arXiv:1207.4059
{[}hep- th{]}{]}; G. Dvali and C. Gomez, JCAP 1401, 023 (2014), {[}arXiv:1312.4795
{[}hep-th{]}{]}.

\bibitem{Dvali:2015} G. Dvali, C. Gomez, R. S. Isermann, D. Lu?st
and S. Stieberger, Nucl. Phys. B 893 (2015),

\bibitem{Casadio:2015} R. Casadio, A. Giugno and A. Orlandi, Phys.
Rev. D 91, no. 12, 124069 (2015), {[}arXiv:1504.05356 {[}gr-qc{]}{]};
R. Casadio, A. Giugno and A. Giusti, Phys. Lett. B 763, 337 (2016),
{[}arXiv:1606.04744 {[}hep-th{]}{]}; R. Casadio, {[}arXiv:2103.14582
{[}gr-qc{]}{]}.

\bibitem{Dvali:2021} G. Dvali, JHEP 03 (2021)

\bibitem{Bose:2021ekc} S.~Bose, A.~Mazumdar and M.~Toro\v{s}, ``Gravitons
in a box,'' Phys. Rev. D \textbf{104} (2021) no.6, 066019 doi:10.1103/PhysRevD.104.066019
{[}arXiv:2104.12793 {[}gr-qc{]}{]}.

\bibitem{Burinskii:2005mm} A.~Burinskii, ``The Dirac-Kerr electron,''
Grav. Cosmol. \textbf{14} (2008), 109-122 doi:10.1134/S0202289308020011
{[}arXiv:hep-th/0507109 {[}hep-th{]}{]}.

\bibitem{Mathur:2005zp} S.~D.~Mathur, ``The Fuzzball proposal
for black holes: An Elementary review,'' Fortsch. Phys. \textbf{53}
(2005), 793-827 doi:10.1002/prop.200410203 {[}arXiv:hep-th/0502050
{[}hep-th{]}{]}; S.~D.~Mathur, ``The Information paradox: A Pedagogical
introduction,'' Class. Quant. Grav. \textbf{26} (2009), 224001 doi:10.1088/0264-9381/26/22/224001
{[}arXiv:0909.1038 {[}hep-th{]}{]}; O.~Lunin and S.~D.~Mathur,
``AdS / CFT duality and the black hole information paradox,'' Nucl.
Phys. B \textbf{623} (2002), 342-394 doi:10.1016/S0550-3213(01)00620-4
{[}arXiv:hep-th/0109154 {[}hep-th{]}{]}.

\bibitem{Mathur:2020ely} S.~D.~Mathur, ``The VECRO hypothesis,''
doi:10.1142/S0218271820300098 {[}arXiv:2001.11057 {[}hep-th{]}{]}.

\bibitem{Mathur:2008kg} S.~D.~Mathur, ``Tunneling into fuzzball
states,'' Gen. Rel. Grav. \textbf{42} (2010), 113-118 doi:10.1007/s10714-009-0837-3
{[}arXiv:0805.3716 {[}hep-th{]}{]}.

\bibitem{Guo:2017jmi} B.~Guo, S.~Hampton and S.~D.~Mathur, ``Can
we observe fuzzballs or firewalls?,'' JHEP \textbf{07} (2018), 162
doi:10.1007/JHEP07(2018)162 {[}arXiv:1711.01617 {[}hep-th{]}{]}.

\bibitem{Maggio:2020jml} E.~Maggio, L.~Buoninfante, A.~Mazumdar
and P.~Pani, ``How does a dark compact object ringdown?,'' Phys.
Rev. D \textbf{102} (2020) no.6, 064053 doi:10.1103/PhysRevD.102.064053
{[}arXiv:2006.14628 {[}gr-qc{]}{]}.

\bibitem{Quantum-optics} C.C. Gerry, and P.L. Knight. Introductory
quantum optics. Cambridge university press; 2005.

\bibitem{Bekenstein} J. D. Bekenstein, Phys. Rev. D 23 (1981), 287

\bibitem{Muck} W. Mück, Eur. Phys. J. C 73 (2013) 2679; W. Mück,
Eur. Phys. J. C (2015) 75:585;

\bibitem{Gockeler:1997dn} M.~Gockeler, R.~Horsley, V.~Linke, P.~E.~L.~Rakow,
G.~Schierholz and H.~Stuben, ``Is there a Landau pole problem in
QED?,'' Phys. Rev. Lett. \textbf{80} (1998), 4119-4122 doi:10.1103/PhysRevLett.80.4119
{[}arXiv:hep-th/9712244 {[}hep-th{]}{]}.

\bibitem{Biswas:2011ar} T.~Biswas, E.~Gerwick, T.~Koivisto and
A.~Mazumdar, ``Towards singularity and ghost free theories of gravity,''
Phys. Rev. Lett. \textbf{108} (2012), 031101 doi:10.1103/PhysRevLett.108.031101
{[}arXiv:1110.5249 {[}gr-qc{]}{]}; T.~Biswas, A.~Mazumdar and W.~Siegel,
``Bouncing universes in string-inspired gravity,'' JCAP \textbf{03}
(2006), 009 doi:10.1088/1475-7516/2006/03/009 {[}arXiv:hep-th/0508194
{[}hep-th{]}{]}.

\bibitem{DHoker:1999bve} E.~D'Hoker, D.~Z.~Freedman, S.~D.~Mathur,
A.~Matusis and L.~Rastelli, 
Nucl. Phys. B \textbf{562} (1999), 330-352 doi:10.1016/S0550-3213(99)00524-6
{[}arXiv:hep-th/9902042 {[}hep-th{]}{]}.

\bibitem{Biswas:2013kla} T.~Biswas, T.~Koivisto and A.~Mazumdar,
``Nonlocal theories of gravity: the flat space propagator,'' {[}arXiv:1302.0532
{[}gr-qc{]}{]}. 

\bibitem{Modesto} E. T. Tomboulis, Superrenormalizable gauge and
gravitational theories,? hep- th/9702146. L. Modesto, Superrenormalizable
Quantum Gravity?, Phys.Rev. D86 (2012), 044005.

\bibitem{Buoninfante:2018xiw} L.~Buoninfante, A.~S.~Koshelev,
G.~Lambiase and A.~Mazumdar, ``Classical properties of non-local,
ghost- and singularity-free gravity,'' JCAP \textbf{09} (2018), 034
doi:10.1088/1475-7516/2018/09/034 {[}arXiv:1802.00399 {[}gr-qc{]}{]}.

\bibitem{Mathur-1} S. D. Mathur, arXiv:0805.3716 {[}hep-th{]}; S.
D. Mathur, Int. J. Mod. Phys. D 18, 2215 (2009) {[}arXiv:0905.4483
{[}hep-th{]}{]}.

\bibitem{Zurek:2021} Kathryn M. Zurek, ``On Vacuum Fluctuations
in Quantum Gravity and Interferometer Arm Fluctuations,'' {[}arXiv:2012.05870
{[}hep-th{]}{]}.

\bibitem{thooft} G.t Hooft, Conf. Proc. C 930308 (1993), 284-296

\bibitem{Susskind} L. Susskind, J. Math. Phys. 36 (1995), 6377-6396

\bibitem{Poisson:2009pwt} E.~Poisson, ``A Relativist's Toolkit:
The Mathematics of Black-Hole Mechanics,'' doi:10.1017/CBO9780511606601

\bibitem{Koshelev:2017bxd} A.~S.~Koshelev and A.~Mazumdar, ``Do
massive compact objects without event horizon exist in infinite derivative
gravity?,'' Phys. Rev. D \textbf{96} (2017) no.8, 084069 doi:10.1103/PhysRevD.96.084069
{[}arXiv:1707.00273 {[}gr-qc{]}{]}; L.~Buoninfante and A.~Mazumdar,
``Nonlocal star as a blackhole mimicker,'' Phys. Rev. D \textbf{100}
(2019) no.2, 024031 doi:10.1103/PhysRevD.100.024031 {[}arXiv:1903.01542
{[}gr-qc{]}{]}.

\bibitem{Cadoni} M. Cadoni, R. Casadio, A. Giusti, W. Mück, M. Tuveri,
Phys. Lett. B 776 (2018) 242, doi:10.1016/j.physletb.2017.11.058;
M. Cadoni, R. Casadio, A. Giusti, M. Tuveri, Phys. Rev. D 97, 044047
(2018) doi:10.1103/PhysRevD.97.044047; M. Cadoni and M. Tuveri, Phys.
Rev. D 100, 024029 (2019), doi:10.1103/PhysRevD.100.024029
\end{thebibliography}

\end{document}